\documentclass{amsart}
\usepackage{amsmath}
\theoremstyle{plain}
\newtheorem{theorem}{Theorem}
\newtheorem*{theorem*}{Theorem}

\newtheorem*{proposition*}{Proposition}

\newtheorem{corollary*}{Corollary}

\theoremstyle{remark}
\newtheorem*{remark*}{Remark}
\theoremstyle{definition}

\numberwithin{equation}{section}

\newcommand\CC{{\mathbf C}}
\newcommand\into{\int_\Omega}
\newcommand\RR{{\mathbf R}}
\newcommand\spr[1]{\langle#1\rangle}

\newcommand\dbar{\overline\partial}
\newcommand\HH{\mathcal H}

\newcommand\Th{T^{(h)}}
\newcommand\CCN{{\CC^N}}
\newcommand\CCNN{{\CC^{N\times N}}}
\newcommand\Tr{\operatorname{Tr}}
\newcommand\diag{\operatorname{diag}}
\newcommand\UN{{U(N)}}
\newcommand\sumr{\sum_{r=0}^\infty}
\newcommand\hol{_{\text{\rm hol}}}
\newcommand\bep{{\text{\boldmath$\epsilon$}}}
\def\Sb#1\endSb{_{\begin{subarray}{c}#1\end{subarray}}}
\newenvironment{roster}{\begin{list}{}{\leftmargin30pt\def
               \makelabel##1{\hss\llap{##1}}}}{\end{list}}
\newcommand\DD{\mathbf D}
\allowdisplaybreaks[4]

\newcommand\bphi{{\text{\boldmath$\phi$}}}
\newcommand\bpsi{{\text{\boldmath$\psi$}}}
\newcommand\bOmega{{\text{\boldmath$\Omega$}}}
\newcommand\bmu{{\text{\boldmath$\mu$}}}
\newcommand\bHH{{\text{\boldmath$\mathfrak H$}}}
\newcommand\CCnNN{{\CC^{n\times N\times N}}}
\newcommand\intob{\int_\bOmega}
\newcommand\bZ{{\mathbf Z}}
\newcommand\D[1]{D^{(#1)}}
\renewcommand\d[1]{d^{(#1)}}
\newcommand\kh{k^{(h)}}
\newcommand\bD{{\mathbf D}}
\newcommand\bd{{\mathbf d}}
\newcommand\Kh{K^{(h)}}
\newcommand\bX{{\mathbf X}}
\newcommand\bY{{\mathbf Y}}
\newcommand\bup{{\text{\boldmath$\upsilon$}}}

\begin{document}

\title[Berezin-Toeplitz quantization]
{A~matrix-valued Berezin-Toeplitz quantization}
\author{S.~Twareque Ali, M.~Engli\v s}
\address{Mathematics Institute, Silesian University at Opava,
Na~Rybn\'{\i}{\v c}ku~1, 74601~Opava, Czech Republic {\rm and}
Mathematics Institute, \v Zitn\'a 25, 11567 Prague 1, Czech Republic}
\email{englis{@}math.cas.cz}
\thanks{The first author would like to acknowledge support from the NSERC,
Canada and FQRNT, Qu\'ebec. Research of the second author was supported by
GA~\v CR grant no.~201/06/0128 and AV~\v CR research plan no.~AV0Z10190503.}
\address{Department of Mathematics and Statistics, Concordia University,
1455~Blvd.~de Maisonneuve~West, Montr\'eal, Qu\'ebec, Canada H3G 1M8}
\email{stali{@}mathstat.concordia.ca}
\begin{abstract}
We~generalize some earlier results on a Berezin-Toeplitz~type 
of quantization on Hilbert spaces built over certain matrix domains.
In~the present, wider setting, the~theory could be applied to systems
possessing several kinematic and internal degrees of freedom. Our~analysis
leads to an identification of those observables, in~this general context, 
which admit a semi-classical limit and those for which no such limit exists.
It turns out that the latter class of observables involve the internal degrees 
of freedom in an intrinsic way. Mathematically, the theory, being a
generalization of the standard Berezin-Toeplitz quantization, points 
the way to applying such a quantization technique to possibly non-commutative
spaces, to~the extent that points in phase space are now replaced by 
$N\times N$ matrices.  \end{abstract}

\maketitle

\section{Introduction}\label{INTR}
Let $\Omega$ be a symplectic manifold, with symplectic form $\omega$,
and~$\HH$ a subspace of $L^2(\Omega,d\mu)$, for~some measure~$\mu$.
For $\phi\in C^\infty(\Omega)$, the~(generalized) \emph{Toeplitz operator}
$T_\phi$ with symbol $\phi$ is the operator on $\HH$ defined~by
\begin{equation}
T_\phi f = P(\phi f),  \qquad f\in\HH,  \label{tTT}
\end{equation}
where $P:L^2(\Omega,d\mu)\to\HH$ is the orthogonal projection. It~is easily
seen that $T_\phi$ is a bounded operator whenever $\phi$ is a bounded function,
and $\|T_\phi\|_{\HH\to\HH}\le\|\phi\|_\infty$, the supremum norm of~$\phi$.

Suppose now that both the measure $\mu$ and the subspace $\HH$ are made to
depend on an additional parameter~$h>0$ (shortly to be interpreted as the
Planck constant), in~such a way that the associated Toeplitz operators
$\Th_\phi$ on $\HH_h$ satisfy, as~$h\searrow0$,
\begin{equation}
\|\Th_\phi\|_{\HH_h\to\HH_h} \to \|\phi\|_\infty,  \label{tTNUL}
\end{equation}
and
\begin{align}
& \|\Th_\phi \Th_\psi - \Th_{\phi\psi} \|_{\HH_h\to\HH_h} \to 0, \label{tTA} \\
& \|\tfrac{2\pi}{ih}[\Th_\phi,\Th_\psi]- \Th_{\{\phi,\psi\}} \|_{\HH_h\to\HH_h}
\to 0  \label{tTB}
\end{align}
(where $\{\cdot,\cdot\}$ is the Poisson bracket with respect to~$\omega$), and,
more generally,
\begin{equation}
\Th_\phi \Th_\psi \approx \sum_{j=0}^\infty h^j \; \Th_{C_j(\phi,\psi)}
\qquad\text{as } h\to0,  \label{tXA}  \end{equation}
for some bilinear differential operators $C_j:C^\infty(\Omega)\times C^\infty
(\Omega)\to C^\infty(\Omega)$, with $C_0(\phi,\psi)=\phi\psi$ and $C_1(\phi,
\psi)-C_1(\psi,\phi)=\frac i{2\pi}\{\phi,\psi\}$. Here the last asymptotic
expansion means, more precisely, that
\begin{equation}  \Big\| \Th_\phi \Th_\psi - \sum_{j=0}^N h^j
\Th_{C_j(\phi,\psi)} \Big\| _{\HH_h\to\HH_h}
= O(h^{N+1}) \quad\text{as } h\searrow0 , \qquad \forall N=0,1,2,\dots.
\label{tTC}  \end{equation}
One then speaks of the \emph{Berezin-Toeplitz quantization.} Indeed, it~is well
known that the recipe
$$ \phi * \psi := \sum_{j=0}^\infty h^j\, C_j(\phi,\psi)  $$
then gives a \emph{star-product} on ~$\Omega$, and (\ref{tTA}), (\ref{tTB})
just amount to its correct semiclassical limit.

The~simplest instance of the above situation is $\Omega=\RR^{2n}\simeq\CC^n$,
with the standard (Euclidean) symplectic structure, and
\begin{equation}  \HH_h = L^2\hol(\Omega,d\mu_h)   \label{tSB}
\end{equation}
the Segal-Bargmann space of all holomorphic functions square-integrable with
respect to the Gaussian measure $d\mu_h(z):=e^{-\|z\|^2/h} (\pi h)^{-n}\,dz$
($dz$~being the Lebesgue measure on~$\CC^n$). As~shown by Coburn~\cite{Cob},
(\ref{tXA})~then holds with
\begin{equation}  C_j(\phi,\psi) = \sum_{|\alpha|=j} \frac1{\alpha!}
\,\partial^\alpha\phi \,\cdot\, \dbar^\alpha\psi.   \label{tCF}
\end{equation}
The resulting star-product coincides, essentially, with the familiar Moyal
product.

Other examples of Berezin-Toeplitz quantization include the unit disc $\DD$
with the Poincar\'e metric, bounded symmetric domains, strictly pseudoconvex
domains with metrics having reasonable boundary behaviour, or,~provided one
allows not only holomorphic functions but also sections of line bundles as
elements of~$\HH_h$, all compact K\"ahler manifolds whose K\"ahler form is
integral. In~all these cases, the choice of the spaces~(\ref{tSB}) which
works are the weighted Bergman spaces $\HH_h = L^2\hol(\Omega,e^{-\Phi/h}
\omega^n)$ (the~subspaces of all holomorphic functions in $L^2(\Omega,e^{-\Phi
/h}\omega^n)$), where $n$ is the complex dimension of~$\Omega$ and $\Phi$ is a
K\"ahler potential for~$\omega$ (so,~for instance, for~the unit disc $\HH_h
=L^2\hol(\DD,\frac{1-h}{\pi h}(1-|z|^2)^{(1/h)-2}\;dz)$). See~\cite{KS},
\cite{BMS} or~\cite{AE1} for the details and further discussion.

Though this seems not to have been recorded explicitly in the literature,
the~whole formalism also extends seamlessly to spaces of \emph{vector-valued}
functions. In~physical terms, this can be interpreted as accommodating the
\emph{internal degrees of freedom} of the quantized system. Namely, replacing
the spaces $\HH_h$ and $L^2(\Omega,d\mu_h)$ by the tensor products $\HH_h
\otimes\CCN$ and $L^2(\Omega,d\mu_h)\otimes\CCN$ (which can be viewed as
spaces of $\CCN$-valued functions on~$\Omega$), one~can define in the same
way the Toeplitz operators~$\Th_\bphi$, where now the symbol $\bphi$ can even
be allowed to be a~($N\times N$)-\emph{matrix}-valued function on~$\Omega$.
It~is a simple matter to check, however, that this Toeplitz operator is~just
the $N\times N$ matrix $[T_{\bphi_{jk}}]_{j,k=1}^N$ of Toeplitz operators
on~$\HH_h$, and it immediately follows that (\ref{tXA}) remains in force
in this vector-valued situation whenever it holds for the scalar-valued~one.
In~particular, for~scalar-valued functions~$\phi$ (i.e.~$\bphi(z)_{jk}=
\delta_{jk}\phi(z)$ for some $\phi:\Omega\to\CC$), one~recovers (\ref{tXA})
completely, with the same cochains~$C_j$.

In~this paper, we work out a formalism, based upon certain spaces of
matrix-valued functions, which could be looked upon, in appropriate 
cases, as~a~possible different approach to the quantization of the
internal degrees of freedom of systems whose kinematics is defined
on complex phase spaces~$\Omega$. Moreover, the more general setting
adopted here, in that points in phase space are replaced by $N \times N$
matrices, could potentially be used to describe systems defined over
non-commutative spaces. 

In~more concrete terms, our~spaces $\bHH_h$ will be suitable subspaces
(actually, rather small~ones, in~terms of codimension) of the spaces
$L^2(\bOmega,d\bmu_h)\otimes\CCN$ of $\CCN$-valued functions 
on~certain domains $\bOmega$ in $\CCnNN$ associated to $\Omega$ in a
natural~way. (Here, as~before, $n$~is the complex dimension of~$\Omega$
and $N$ is related to the number of internal degrees of freedom.)
The~Toeplitz operators are again defined by~the formula~(\ref{tTT}), 
only with $P$ replaced by the orthogonal projection onto~$\bHH_h$, 
and the symbol $\bphi$ can now be allowed to be a~$\CCNN$-valued function
on~$\bOmega$. Finally, there exists a canonical unitary isomorphism
$\iota:\bHH_h\to\HH_h\otimes\CCN$ (with $\HH_h$ as in (\ref{tSB})). 
This means that the quantum system defined on $\bHH_h$ can be thought 
of as one possessing $N$ internal degrees of freedom and moving on 
the phase space $\Omega$.

The~following facts then emerge from our analysis.
\begin{roster}
\item[(a)] To~any function $\phi$ on $\Omega$ one can associate, in~a
canonical~way, a~function $\bphi$ on~$\bOmega$. (For~reasons which will
become apparent later, functions $\bphi$ that arise in this way will be
called \emph{spectral functions.}) For~any two functions $\bphi,\bpsi$ of
this form, the~corresponding Toeplitz operators turn out to be unitarily
equivalent via $\iota$ to $\Th_\phi\otimes I$ and $\Th_\psi\otimes I$,
respectively, acting on~$\HH_h\otimes\CCN$. Consequently, (\ref{tXA})~must
hold (with the same cochains~$C_j$), and, in~this sense, our~quantization
contains the original scalar-valued Berezin-Toeplitz quantization, as~well
as its vector-valued analogue obtained by tensoring with~$\CCN$ (and~using
only scalar-valued symbols~$\bphi$), mentioned~above.
\item[(b)] Let~the unitary group $\UN$ of order $N$ act on $\CCnNN$~by
$$\quad  \bZ^U=(U^*Z_1U,U^*Z_2U,\dots,U^*Z_nU) \quad \text{with}\quad
\bZ=(Z_1,\dots,Z_n)\in\CCnNN.  $$
The~domain $\bOmega$ is invariant under this action, and functions $\bphi$
satisfying $\bphi(\bZ^U)=U^*\bphi(\bZ)U$ $\forall U\in\UN$ will be called
\emph{$U$-invariant.} All~spectral functions are $U$-invariant, but~not vice
versa. It~is~then the case that for any $U$-invariant function~$\bphi$,
the~Toeplitz operator $\Th_\bphi$ is unitarily equivalent via $\iota$ to 
the operator $\Th_{\pi_h\bphi}\otimes I$ on~$\HH_h\otimes\CCN$, where
$\pi_h\bphi\in C^\infty(\Omega)$ is a certain ``average'' of~$\bphi$ 
over the internal variables (reminiscent of ``spin averaging'' in quantum
mechanical scattering theory). Not~surprisingly, the~operator $\pi_h$ behaves
nicely as~$h\to0$; owing to this, for~any two $U$-invariant functions $\bphi,
\bpsi$ one obtains a semiclassical expansion of the product $\Th_\bphi
\Th_\bpsi$ of~the~form
$$ \Th_\bphi\Th_\bpsi \approx \sumr h^r \Th_{\bup_r}  $$
for some uniquely determined \emph{spectral} functions $\bup_r$.
Thus, in~this sense, the~internal degrees of freedom disappear in the
semiclassical limit, as~they should. Using the isomorphism~$\iota$ and the
facts mentioned in~(a), this can also be recast into the language of the
traditional vector-valued quantization discussed before; note,~however, 
that now we are able to quantize not only the scalar-valued functions (which 
we have seen in (a) to correspond to the spectral functions on~$\bOmega$),
but~a \emph{much wider} class of observables corresponding to $U$-invariant
functions.  
\item[(c)] Finally, for~completely general functions $\bphi,\bpsi\in C^\infty
(\bOmega)$, the~semiclassical expansion of the product $\Th_\bphi \Th_\bpsi$
in~the usual sense (i.e.,~in~the sense of~(\ref{tXA})) does not exist.
(There may be~one, but~the cochains $C_j$ are then no longer uniquely
determined unless one requires that their values always be spectral functions,
and~then they are no longer local (i.e.~differential) operators, but~rather 
involve some kind of averaging over a sort of $\UN$-orbit~of~$\bZ$.)
Consequently, such functions lead to quantum observables that have no 
classical counterparts. The~following situation is thus seen to emerge: 
while~the Toeplitz operator corresponding a general function $\bphi$ 
could be a legitimate quantum observable, only those functions which 
are $U$-invariant, and consequently involve the internal degrees of 
freedom only in a ``controlled''~way, admit a semi-classical limit. 
In~other words, only observables kinematically related to the phase space 
have semi-classical limits. (Note that even in the case where internal 
degrees of freedom are absent, i.e., $N=1$, the model of quantum mechanics
being used here is one where the wave functions are defined on phase space 
and not on configuration space.) 
\end{roster}

The~whole approach is applicable to any phase space $\Omega\subset\CC^n$
admitting the ordinary (i.e.~scalar-valued) Berezin-Toeplitz quantization.
At~the moment, we~do not know how to extend it from domains in $\CC^n$ to
manifolds.

For~the simplest case of $\Omega=\CC$, corresponding to a free particle on~the
real~line, the~results above have been obtained in~\cite{AE2}. For~the reader's
convenience, we~review, in~Section~\ref{ZWEI} below, the necessary material
from that paper (without proofs), as~well as from its precursor \cite{AEG},
where spaces of matrix-valued functions of this type were first introduced.
The~quantization procedure is spelled out in Section~\ref{FUER}. Hidden under
surface in all these developments are also certain vector- and matrix-valued
analogues of some reproducing kernels and coherent states; these in fact make
sense in several more general situations as well (even though the quantization
procedure may not lead to physically meaningful theories). We describe these 
in the last Section~\ref{DREI}. 

A~word of clarification is, perhaps, in order at this juncture. We~are not
suggesting here that the current formalism be used to replace the traditional
quantum mechanical setup for describing systems with internal degrees of
freedom. As~far as traditional quantum mechanics is concerned, the~present
formalism, with wave functions described over matrix domains is an interesting
alternative~to~it. Besides being well-adapted to studying the semi-classical
limit, the~present formalism can also be easily employed to build ``quantum
systems'' which show no limiting semi-classical behaviour at all! One~might
venture a guess that such quantum systems (which even admit a proper
probability interpretation on ``phase space'') could point to some
underlying non-commutative geometry.

\medskip

\section{The case of the complex plane}   \label{ZWEI}
For~the reader's convenience, we~briefly review here the salient facts
from~\cite{AE2} and~\cite{AEG}, which correspond to the simplest case of
the quantization on~$\Omega=\CC$.

Consider the domain $\bOmega=\{Z\in\CCNN:\;Z^*Z=ZZ^*\}$ of all normal matrices
in~$\CCNN$. By~the spectral theorem, any $Z\in\bOmega$ can be written in
the~form
\begin{equation}  Z= U^* D U,  \label{tUDU}  \end{equation}
with $U\in\UN$ unitary and $D$ diagonal; $D$~is determined by $Z$ uniquely up
to permutation of the diagonal elements, and if the latter are all distinct and
their order has been fixed in some way, then $U$ is unique up to left
multiplication by a diagonal matrix with unimodular elements. Consequently,
there exists a unique measure $d\bmu_h(Z)$ on $\bOmega$ such that
\begin{equation}  \intob f(Z) \,d\bmu_h(Z) = (\pi h)^{-N} \int_\UN \int_\CCN
f(U^* DU) \, e^{-\|D\|^2/h} \,dU \, dD  \qquad \forall f,   \label{dUF}
\end{equation}
where $dU$ is the normalized Haar measure on~$\UN$, $dD$~is the Lebesgue
measure on~$\CCN$, where we are identifying the diagonal matrix $D=\diag
(d_1,\dots,d_N)$ with the vector $d=(d_1,\dots,d_N)\in\CCN$, and $\|D\|^2
=\|d\|^2:=|d_1|^2+\dots+|d_N|^2$. It~can be shown \cite{AEG} that
\begin{equation}  \intob Z^{*j} Z^k \, d\bmu_h(Z) = \delta_{jk} k! h^k I,
\label{tUOG}  \end{equation}
so~that the elements
\begin{equation}  \frac {Z^k \chi_j}{\sqrt{k! h^k}}, \qquad j=1,\dots,N,
\ k=0,1,2,\dots, \label{tZK}  \end{equation}
where $\chi_1,\dots,\chi_N$ is the standard basis of~$\CCN$, are~orthonormal
in $L^2(\bOmega,d\bmu_h)\otimes\CCN$. Let~$\bHH_h$ be the subspace spanned
by these functions.

In~analogy with the scalar-valued situation, we~next define for any $\bphi\in
C^\infty(\Omega)\otimes\CCNN$ the Toeplitz operator $T_\bphi$ on $\bHH_h$ by
the recipe
$$ \Th_\bphi f = P_h (\bphi f),  $$
where $P_h: L^2(\bOmega,d\bmu_h)\to\bHH_h$ is the orthogonal projection.
Note~that the last formula implies that $\|\Th_\bphi\|_{\bHH_h\to\bHH_h}
\le\|\bphi\|_\infty:=\sup_{X\in\bOmega}\|\bphi(X)\|_{\CCN\to\CCN}$.

A~$\CCNN$-valued function $\bphi(Z)$ of $Z\in\bOmega$ will be called
\emph{spectral} if it is a function of $Z$ in the sense of the Spectral
Theorem for matrices: that~is, if~there exists a function $\phi:\CC\to\CC$
such that $\bphi=\phi^\#$, where
\begin{equation}  \phi^\#(Z) := U \cdot \diag_j(\phi(d_j)) \cdot U^*
\qquad\text{if}\quad Z=U\cdot\diag_j(d_j)\cdot U^*.  \label{tGH}
\end{equation}
Further, as~was already mentioned in the Introduction, the~function $\bphi$
will be called \emph{$U$-invariant}~if
\begin{equation}  \phi(U^*ZU) = U^*\,\phi(Z) \,U \qquad\forall U\in\UN
 \;\forall Z\in\bOmega.  \label{tUI}   \end{equation}
Clearly, a~spectral function is $U$-invariant, but not vice versa: an~example
is the function $\bphi(Z)=|\!\det Z|^2 I$.

The~following results have been established in~\cite{AE2}.

\begin{proposition*} {\rm(\cite{AE2}, Proposition~12)} A~function $\bphi$ is
$U$-invariant if and only if there exists a function $\phi(d_1;d_2,\dots,d_N)$
from $\CC\times\CC^{N-1}$ into~$\CC$, symmetric in the last $N-1$ variables
$d_2,\dots,d_N$, such that $\bphi=\phi^\#$, where
\begin{equation}  \phi^\# (U^*DU) := U^*\cdot \diag_j(\phi(d_j;d_1,\dots,\hat
d_j,\dots,d_N))  \cdot U.  \label{tUJ}  \end{equation}
The function $\phi$ is uniquely determined by~$\bphi$.

Further, $\bphi$ is spectral if and only if $\phi$ depends only on the first
variable, i.e.~if and only if $\phi(d_1;d_2,\dots,d_N)=\phi(d_1;0,\dots,0)$.
\end{proposition*}

(Here we are using the notation $\phi^\#$ both in the sense of (\ref{tUJ})
and~(\ref{tGH}), but there is no danger of confusion.)

\begin{theorem*} {\rm(\cite{AE2}, Theorem~10)} If~$\bphi=\phi^\#$ and
$\bpsi=\psi^\#$ are two smooth spectral functions on~$\bOmega$, then there
exist unique spectral functions $\bup_r$, $r=0,1,2,\dots$, such~that
$$ \Th_\bphi\Th_\bpsi \approx \sumr \,h^r\;\Th_{\bup_r}
\qquad\text{as }h\to0  $$
in~the sense of operator norms $($i.e.~as~in~$(\ref{tTC}))$. In~fact,
$$ \bup_r = C_r(f,g)^\# ,  $$
where
\begin{equation} C_r(f,g)=\frac1{r!}\,\partial^r f\cdot\dbar^r g
\label{tCR}  \end{equation}
are the operators $(\ref{tCF})$ for $n=1$.  \end{theorem*}

\begin{theorem*} {\rm(\cite{AE2}, Theorem~16)} For~a~function $f$ on $\CCN$
and $h>0$, let $\pi_h f$ be the function on~$\CC$ defined~by
$$ \pi_h f(z_1) := \int_{\CC^{N-1}} f(z_1,z_2,\dots,z_N) \;
e^{-(|z_2|^2+\dots+|z_N|^2)/h} \; \frac{dz_2\dots dz_N}{(\pi h)^{N-1}}.  $$
Let $\bphi=\phi^\#$, $\bpsi=\psi^\#$ be smooth $U$-invariant functions
on~$\bOmega$ such that the partial derivatives of $\phi$ and $\psi$ of all
orders are bounded, and let $C_r$ be the bidifferential
operators~$(\ref{tCR})$. Then
\begin{equation}  \begin{aligned}
\Th_\bphi\Th_\bpsi &\approx \sumr\;h^r \,\Th_{C_r(\pi_h \phi,\pi_h \psi)^\#} \\
&\approx \sumr h^r \; \Th_{\upsilon_r^\#},   \end{aligned}  \label{tRR}
\end{equation}
in the sense of operator norms, where
$$ \upsilon_r = \sum\Sb j,k,l\ge0,\\ j+k+l=r\endSb \frac1{j!k!l!} \, \partial^l
(\Delta^{\prime j}\phi)^\flat \cdot \dbar^l(\Delta^{\prime k}\psi)^\flat ,  $$
where $\Delta'$ denotes the Laplacian with respect to the last $N-1$
variables $z_2,\dots,z_N$, and $f^\flat(z):=f(z;0,\dots,0)$.   \end{theorem*}

Finally, for~functions which are not $U$-invariant, things seem to go
wrong regarding quantization: namely, there is evidence that in general the
semiclassical expansion of the form (\ref{tXA}) either does not exist, or~if
it exists then the cochains $C_j$ have rather pathological properties
(for~instance, are~not local operators --- the~value of $C_j(\bphi,\bpsi)$
at~a~point $Z$ need not depend only on the jets of $\bphi$ and $\bpsi$ at~$Z$).
In~more detail: first of~all, there exist functions $\bphi$ (even very nice and
$U$-invariant ones --- for~instance, $\phi(Z)=|\!\det(Z)|^2 e^{-\Tr(Z^*Z)}I$)
for~which $\|\Th_\bphi\|\to0$ as~$h\searrow0$; as~a~result, the~cochains $C_j$
in (\ref{tXA}) have no chance of being uniquely determined, unless they are
subjected to some additional condition. The~only such condition which gives
the right answer for spectral functions seems to be that $C_j$ take values
in spectral functions; let~us therefore assume that this is the case. Second,
there exist families of elements $\kh_{Z,\chi}\in\bHH_h$, labelled by~$Z\in
\bOmega$ and $\chi\in\CCN$ (interpretable as normalized reproducing kernels,
or~vector \emph{coherent states} --- see~Section~\ref{DREI} below for more
information), such~that as~$h\nearrow0$, there are asymptotic expansions
\begin{align*}
\spr{\Th_\bphi \kh_{Z,\chi},\kh_{Z,\eta}} &\approx
\sumr h^r \; \eta^* l_r[\bphi](Z) \chi ,  \\
\spr{\Th_\bphi\Th_\bpsi \kh_{Z,\chi},\kh_{Z,\eta}} &\approx
\sumr h^r \; \eta^* m_r[\bphi,\bpsi](Z) \chi ,  \end{align*}
for some $\CCNN$-valued functions $l_r[\bphi]$ and $m_r[\bphi,\bpsi]$
on~$\bOmega$, $r=0,1,2,\dots$. If~(\ref{tXA}) holds, then we~must therefore
have $m_0[\bphi,\bpsi]=l_0[C_0(\bphi,\bpsi)]$. For~spectral functions,
$l_0$~turns out to be just the identity operator; since we have agreed that
$C_j$ take values in spectral functions, it~follows that $C_0(\bphi,\bpsi)=
m_0[\bphi,\bpsi]$. Now~computations show that $m_0[\bphi,\bpsi](Z)$ is given
by a rather complicated expression involving integration over the whole orbit
$\{U^*(d_j\spr{\cdot,\chi_j}\chi_j)U\}$ of the spectral projections $d_j\spr
{\cdot,\chi_j}\chi_j$, $j=1,\dots,N$, of~$Z$ under the unitary group~$\UN$.
The~reader is referred to Sections~4--6 of \cite{AE2} for the full story.

The~appearance of $\phi^\flat$ and~$\psi^\flat$, and not $\phi$ and~$\psi$,
in~(\ref{tRR}) means that the $\CC^{N-1}$ part of $\phi$ disappears in the
semiclassical limit $h\to0$, and only the projection~$\phi^\flat$, which
 lives on~$\CC$, survives; that~is, only the ``spectral component'' of the
corresponding $U$-invariant function $\bphi$ on~$\bOmega$. As mentioned before,
all this means that we are dealing here with a quantum system which has $N$
internal degrees of freedom, and that the full set of quantum observables of
this system includes those which do not have classical counterparts, while
even for those having the classical counterparts, the~internal degrees of
freedom --- being purely quantum in this case --- do~not survive in the
semi-classical limit.

\section{General domains} \label{FUER}
We~proceed to describe how the~spaces from the preceding section can be adapted
from the complex plane to any phase-space $\Omega\subset\CC^n$ admitting the
ordinary (scalar-valued) Berezin-Toeplitz quantization.

The~appropriate matrix domain~is
\begin{align*}
\bOmega = & \{\bZ=(Z_1,Z_2,\dots,Z_n)\in\CCnNN: \\
& \qquad \qquad \; Z_j Z^*_k=Z^*_k Z_j \, \forall j,k=1,\dots,n,
\text{ and } \sigma(\bZ)\subset\Omega \} ,
\end{align*}
i.e.~the set of all commuting $n$-tuples of normal $N\times N$ matrices whose
joint spectrum $\sigma(\bZ)$ is contained in~$\Omega$. In~other words, this
means that in the decomposition (\ref{tUDU}) for the entries~$Z_j$, the~unitary
parts will be the same for all~$j$:
\begin{align*}
& \bZ\in\bOmega \iff \bZ=(U^*\D1 U,\dots,U^*\D n U), \\
& \hskip7em \text{with } U\in\UN \text{ and $\D1,\dots,\D n$ diagonal,}
\end{align*}
and, if~we denote the diagonal entries of the matrices $\D j$ by $\d j_k$
($j=1,\dots,n$, $k=1,\dots,N$),
$$ (\d 1_k,\dots,\d n_k) \in\Omega, \qquad \forall k=1,\dots,N.  $$
With this notation, we~define the measure on $\bOmega$~by
$$ d\bmu_h(\bZ) = \frac1{\mu_h(\Omega)^{N-1}} \,
dU \, \prod_{k=1}^N d\mu_h(\bd_k),  $$
where $\bd_k:=(\d 1_k,\dots,\d n_k)$. We~will also sometimes use the shorthand
$$ \diag(\bd_1,\dots,\bd_n)  $$
to~denote the $n$-tuple of diagonal matrices $(\D1,\dots,\D n)\in\bOmega$.

It~remains to define the spaces $\bHH_h$. Observe that since the $Z_j$ commute and
$\sigma(\bZ)\in\Omega$, the~spectral theorem implies that, for~any function
$f:\Omega\to\CC$, we~can form the matrix $f(Z_1,\dots,Z_n)=:f^\#(\bZ)\in\CCNN$:
specifically,
\begin{equation} \begin{aligned}
f^\#(\bZ) &= U^* \diag_k (f(\d 1_k,\dots,\d n_k)) \, U
&= U^* \diag_k (f(\bd_k)) \, U.   \end{aligned}  \label{dGH}  \end{equation}
Functions on $\bOmega$ of this form will be called \emph{spectral functions.}
We~now define spaces $\bHH_h\subset L^2(\bOmega,d\bmu_h)\otimes\CCN$~as
$$ \bHH_h = \operatorname{span} \{ f^\#(\bZ) \chi : \; f
\in L^2\hol(\Omega,d\mu_h), \; \chi\in\CCN \}.  $$

Finally, recall also from the Introduction that a function $\bphi:\bOmega\to
\CCNN$ is called \emph{$U$-invariant}~if $\bphi(\bZ^U)=U^*\bphi(\bZ)U$ for all
$\bZ\in\bOmega$ and $U\in\UN$, where $\bZ^U=(U^*Z_1U,U^*Z_2U,\dots,U^*Z_nU)$.
(Clearly, this reduces to the definition from Section~\ref{ZWEI} if~$n=1$.)

Our~main result is the following.

\begin{theorem} \label{vBU} {\rm(i)} The~mapping
\begin{equation} \iota: f(z) \otimes \chi \mapsto f^\#(\bZ) \chi   \label{tIO}
\end{equation}
is~a unitary isomorphism of $L^2\hol(\Omega,\mu_h)\otimes\CCN$ onto~$\bHH_h$.

{\rm(ii)} Under this isomorphism, the~Toeplitz operator~$\Th_\bphi$, for~a
spectral function~$\bphi=\phi^\#$, corresponds to the tensor product $\Th_\phi
\otimes I$ of the scalar Toeplitz operator $\Th_\phi$ on $L^2\hol(\Omega,d\mu_h)$
with the identity operator~on~$\CCN$. $($In~other words --- to~the Toeplitz
operator $\Th_{\phi I}$ on $L^2\hol(\Omega,d\mu_h)\otimes\CCN$ with scalar
matrix-valued symbol $\phi I$ discussed in the second paragraph after
$(\ref{tCF})$ in the Introduction.$)$

{\rm(iii)} Consequently, if~$\bphi=\phi^\#$ and $\bpsi=\psi^\#$ are two smooth
spectral functions on~$\bOmega$, then there exist unique spectral functions
$\bup_r$, $r=0,1,2,\dots$, such~that
$$ \Th_\bphi\Th_\bpsi \approx \sumr \,h^r\;\Th_{\bup_r}
\qquad\text{as }h\to0  $$
in~the sense of operator norms $($i.e.~as~in~$(\ref{tTC}))$. In~fact,
$$ \bup_r = C_r(f,g)^\# ,  $$
where $C_r$ are the cochains $(\ref{tXA})$ from the ordinary
$($i.e.~scalar-valued$)$ Berezin-Toeplitz quantization on~$\Omega$.

{\rm(iv)} A~function $\bphi:\bOmega\to\CCNN$ is $U$-invariant if and only
if there exists a function $\phi(\bd_1;\bd_2,\dots,\bd_N)$ from $\Omega\times
\Omega^{N-1}$ into~$\CC$, symmetric in the last $N-1$ variables $\bd_2,\dots,
\bd_N$, such that $\bphi=\phi^\#$, where
\begin{equation}
\phi^\#(\bZ) := U^* (\diag_k(\phi(\bd_k;\bd_1,\dots,
\hat\bd_k,\dots,\bd_N))) U.  \label{dUJ}  \end{equation}
The~function $\phi$ is uniquely determined by~$\bphi$, and $\bphi$ is spectral
if and only if $\phi$ depends only on the first variable, i.e.~if and only if
$\phi(\bd_1;\bd_2,\dots,\bd_N)=\phi(\bd_1)$.

{\rm(v)} For~a $U$-invariant function $\bphi=\phi^\#$, the Toeplitz operator
$\Th_\bphi$ corresponds, under the isomorphism~$(\ref{tIO})$, to~the tensor
product $\Th_{\pi_h\phi}\otimes I$, where
$$ \pi_h f(z_1) := \frac1{\mu_h(\Omega)^{N-1}} \int_{\Omega^{N-1}}
f(z_1;z_2,\dots,z_N) \, \prod_{j=2}^N d\mu_h(z_j).   $$

{\rm(vi)} Consequently, for any two smooth $U$-invariant functions $\bphi=\phi
^\#$, $\bpsi=\psi^\#$ on~$\bOmega$ such that the ordinary $($scalar-valued$)$
Berezin-Toeplitz quantization on~$\Omega$ is applicable to~$\phi$ and~$\psi$,
\begin{equation}  \Th_\bphi \Th_\bpsi \approx \sumr h^r \;
\Th_{C_r(\pi_h\phi,\pi_h\psi)^\#} \label{dUC}  \end{equation}
in~the sense of operator norms, where $C_r$ are the cochains $(\ref{tXA})$ from
the ordinary Berezin-Toeplitz quantization on~$\Omega$.

{\rm(vii)} Finally, if,~in~addition, $\Omega$~is one of the domains mentioned
in the paragraph after~$(\ref{tCF})$ in the Introduction $($examples of domains
on which the scalar-valued Berezin-Toeplitz quantization is currently known to
work$)$, and $\phi$ and $\psi$ have compact support, then $(\ref{dUC})$ can
be converted into an asymptotic expansion in powers of~$h$, i.e.~there exist
uniquely determined spectral functions $\bup_r$, $r=0,1,\dots$,
such~that
\begin{equation} \Th_\bphi\Th_\bpsi \approx \sumr h^r \; \Th_{\upsilon_r^\#}
\label{tRD} \end{equation}
in the sense of operator norms.    \end{theorem}

The~hypotheses in the part~(vii) are made only for technical reasons, and~can
probably be weakened or dropped altogether.

\begin{proof} (i)~For any $f,g\in L^2\hol(\Omega,d\mu_h)$ and $\chi,\eta\in
\CCN$, we~have
\begin{align*}
&\spr{f^\#(\bZ)\chi,g^\#(\bZ)\eta}
= \intob \eta^* g^\#(\bZ)^* f^\#(\bZ) \chi \,d\bmu_h(\bZ) \\
&\qquad= \mu_h(\Omega)^{1-N} \; \int_\UN \int_{\Omega^N} \eta^* U^*
\diag_k(\overline{g(\bd_k)}) \diag_k(f(\bd_k)) U \chi
\,dU \, \prod_j d\mu_h(\bd_j) . \end{align*}
Since, for any matrix~$X$,
\begin{equation}  \int_\UN U^* X U \,dU = \frac{\Tr(X)}N \, I,  \label{tUU}
\end{equation}
we~can continue the computation~by
\begin{align*}
&= \frac1N \; \mu_h(\Omega)^{1-N} \; \eta^*\chi \int_{\Omega^N}
\sum_k \overline{g(\bd_k)} f(\bd_k) \, \prod_j d\mu_h(\bd_j)  \\
&= \frac1N \; \mu_h(\Omega)^{1-N} \; \eta^*\chi \sum_k
\Big(\into\,d\mu_h\Big)^{N-1} \into \overline{g(\bd_k)} f(\bd_k)
\, d\mu_h(\bd_k) \\
&= \frac1N \; \eta^*\chi N \spr{f,g} \\
&= \spr{\chi,\eta} \spr{f,g} ,   \end{align*}
and the claim follows.

(ii)~For $f,g\in L^2\hol(\Omega,d\mu_h)$, $\chi,\eta\in\CCN$ and any function
$\phi$~on~$\Omega$, we~have by a similar computation as~in~(i),
\begin{align*}
&\spr{\Th_{\phi^\#} f^\# \chi,g^\# \eta}
= \intob \eta^* g^\#(\bZ)^* \phi^\#(\bZ) f^\#(\bZ) \chi \,d\bmu_h(\bZ) \\
&\qquad= \mu_h(\Omega)^{1-N} \; \int_\UN \int_{\Omega^N} \eta^* U^*
\diag_k(\overline{g(\bd_k)}) \diag{\phi(\bd_k)} \\
&\hskip6em\vphantom{\int} \diag_k(f(\bd_k)) U
\chi \,dU\, \prod_j d\mu_h(\bd_j)  \\
&\qquad= \frac1N \; \mu_h(\Omega)^{1-N} \; \eta^*\chi \int_{\Omega^N}
\sum_k \overline{g(\bd_k)} \phi(\bd_k) f(\bd_k) \, \prod_j d\mu_h(\bd_j)  \\
&\qquad= \frac1N \; \mu_h(\Omega)^{1-N} \; \eta^*\chi \sum_k
\Big(\into\,d\mu_h\Big)^{N-1} \into \overline{g(\bd_k)} \phi(\bd_k) f(\bd_k)
\, d\mu_h(\bd_k) \\
&\qquad= \frac1N \; \eta^*\chi N \spr{\phi f,g} \\
&\qquad= \spr{\chi,\eta} \spr{\Th_\phi f,g} .   \end{align*}

(iii)~follows immediately from (ii) and the ordinary Berezin-Toeplitz
quantization on~$\Omega$, upon tensoring with~$\CCN$.

(iv)~Let $\bphi:\bOmega\to\CCNN$ be a $U$-invariant function, and let
$\bD=(\D1,\dots,\D n)$ be an element of $\bOmega$ whose entries are
diagonal matrices. For any complex numbers $\epsilon_1,\dots,\epsilon_N$
of modulus~one, consider the matrix $\bep=\diag(\epsilon_1,\dots,\epsilon_N)$.
Then $\bep\in\UN$ and $\bep D\bep^*=D$ for any diagonal matrix~$D$, whence
$\bD^\bep=\bD$; thus by~the $U$-invariance condition,
$$ \bphi(\bD) = \bep^* \,\bphi(\bD) \, \bep \qquad \forall
\epsilon_1,\dots,\epsilon_N\in\mathbf T.  $$
Consequently, $\bphi(\bD)$ is also a diagonal matrix. Define the functions
$f_1,\dots,f_N:\Omega^N\to\CC$~by
$$  f_j(\bd_1;\bd_2,\dots,\bd_N) := (\bphi(\bD))_{jj}
\qquad  \text{where } \bD=\diag(\bd_1,\dots,\bd_N).    $$
For any permutation $\sigma$ of the set $\{1,2,\dots,N\}$, let $F_\sigma$
denote the permutation matrix $[F_\sigma]_{jk}=\delta_{\sigma(j),k}$. Then
$F_\sigma\in\UN$ and
$$ F_\sigma \bD F_\sigma^* = \diag(\bd_{\sigma(1)},\dots,\bd_{\sigma(N)})
\qquad \text{if } \bD=\diag(\bd_1,\dots,\bd_N).  $$
Thus by the $U$-invariance condition again
$$ f_{\sigma(j)}(\bd_1;\bd_2,\dots,\bd_N) = f_j(\bd_{\sigma(1)};
\bd_{\sigma(2)},\dots,\bd_{\sigma(N)}).  $$
It~follows that $f_j$ is symmetric with respect to the last $N-1$ variables
$\bd_1,\dots,\hat\bd_j,\allowbreak\dots,\bd_N$ and $\bphi=f^\#$ for $f=f_1$.

Conversely, it~is easily seen that any function of the form (\ref{dUJ}) is
$U$-invariant, and $f^\#=g^\#\iff f=g$.

Finally, the assertion concerning spectral functions is immediate upon
comparing (\ref{dUJ}) and~(\ref{dGH}).

(v)~Using~(\ref{dUJ}), the~assertion (v) now follows by a similar computation
as in the proofs~of~(i)~and~(ii): for~$f,g\in L^2\hol(\Omega,d\mu_hh)$, $\chi,
\eta\in\CCN$ and any $\phi:\Omega^N\to\CC$ as~in~(iv), we~have
\begin{align*}
&\spr{\Th_{\phi^\#} f^\# \chi,g^\# \eta}
= \intob \eta^* g^\#(\bZ)^* \phi^\#(\bZ) f^\#(\bZ) \chi \,d\bmu_h(\bZ) \\
&= \mu_h(\Omega)^{1-N} \; \int_\UN \int_{\Omega^N} \eta^* U^*
\diag_k(\overline{g(\bd_k)}) \\
&\hskip6em\vphantom{\int}
(\diag_k(\phi(\bd_k;\bd_1,\dots,\hat\bd_k,\dots,\bd_N))) \diag_k(f(\bd_k)) U
\chi \,dU\, \prod_j d\mu_h(\bd_j)  \\
&= \frac1N \; \mu_h(\Omega)^{1-N} \; \eta^*\chi \int_{\Omega^N}
\sum_k \overline{g(\bd_k)} \phi(\bd_k;\bd_1,\dots,\hat\bd_k,\dots,\bd_N)
f(\bd_k) \, \prod_j d\mu_h(\bd_j)  \\
&= \mu_h(\Omega)^{1-N} \; \eta^*\chi \int_{\Omega^N} \overline{g(\bd_1)}
\phi(\bd_1;\bd_2,\dots,\bd_N) f(\bd_1) \, \prod_j d\mu_h(\bd_j) \\
&= \mu_h(\Omega)^{1-N} \; \eta^*\chi \into \overline{g(\bd_1)}
\Big( \int_{\Omega^{N-1}} \phi(\bd_1;\bd_2,\dots,\bd_N) \, \prod_{j=2}^N
d\mu_h(\bd_j) \Big) \, f(\bd_1) \, d\mu_h(\bd_1) \\
&= \eta^*\chi \into \overline{g(\bd_1)} \pi_h\phi(\bd_1) f(\bd_1)
\, d\mu_h(\bd_1) \\
&= \eta^*\chi \spr{(\pi_h\phi) f,g} \\
&= \spr{\chi,\eta} \spr{\Th_{\pi_h\phi} f,g} .   \end{align*}

(vi)~With (v) in hands, we~obtain from the ordinary Berezin-Toeplitz
quantization on~$\Omega$, for~any $\phi,\psi:\Omega^N\to\CC$ as~in~(iv),
\begin{align*}
\Th_{\phi^\#}\Th_{\psi^\#} &\cong \Th_{\pi_h\phi} \Th_{\pi_h\psi} \otimes I\\
&\approx \sumr h^r \, \Th_{C_r(\pi_h\phi,\pi_h\psi)} \otimes I \\
&\cong \sumr h^r \, \Th_{C_r(\pi_h\phi,\pi_h\psi)^\#} ,  \end{align*}
in~the sense of operator norms (the~last isomorphism being the one for
spectral functions from part~(i)), which proves~(vi).

(vii)~Finally, to~convert the last expansion into one of the form~(\ref{tXA})
(i.e.~in powers of~$h$), we~only need to exhibit a uniform asymptotic expansion
for~$\pi_h\phi$, i.e.~show that
\begin{equation}  \pi_h \phi \approx \sumr h^r \, L_r \phi ,  \label{dUA}
\end{equation}
in~the sense of norms in~$L^\infty(\Omega)$, for~some linear operators~$L_r$
acting from functions on~$\Omega^N$ into functions on~$\Omega$. Indeed, since
$C_r$ are bidifferential operators with smooth coefficients and $\phi,\psi$ are
assumed to have compact support, it~will then follow that
$$ C_r(\pi_h\phi,\pi_h\psi)\approx\sum_{j,k\ge0} h^{j+k}C_r(L_k\phi,L_m\psi) $$
in~the sense of $L^\infty$ norms, and~in~view of the inequality $\|\Th_\upsilon
\|\le\|\upsilon\|_\infty$, we~can ``apply~$\Th$'' to~both sides.

In~order to prove~(\ref{dUA}), it~suffices in turn to show that there is an
expansion of that form~for
\begin{equation} h^k \mu_h(\Omega)^{N-1} \pi_h\phi(z) = h^k \int_{\Omega^{N-1}}
\phi(z;z_2,\dots,z_N) \,d\mu_h(z_2)\,\dots\,d\mu_h(z_n),  \label{dUB}
\end{equation}
for some $k\ge0$, with leading coefficient which is positive on $\Omega$ when
$\phi$ is identically~1. Indeed, specializing this to $\phi$ the constant one
and dividing the two expansions gives~(\ref{dUA}). (The~leading coefficient is
needed to make sure that we are not dividing by zero.)

Finally, for~the situations where the ordinary Berezin-Toeplitz quantization
is nowadays known to~work (as~summarized in the paragraph following (\ref{tCF})
in the Introduction), the~measures $\mu_h$ are taken to be $e^{-\Phi/h}d\mu$,
where $\Phi$ is a real-valued potential for the K\"ahler form $\omega$ and
$d\mu(z)=\omega(z)^n=\det[\partial\dbar\Phi(z)]\,dz=:g(z)\,dz$ is the Liouville
measure. However, in~that case the right-hand side of (\ref{dUB}) reduces~to
$$ \int_{\Omega^{N-1}} \phi(z;z_2,\dots,z_N) \; e^{-(\Phi(z_2)+\dots+
\Phi(z_N))/h} \; g(z_2) \dots g(z_N) \; dz_2\dots dz_N,  $$
which has an asymptotic expansion of the desired form by the usual stationary
phase method (or,~rather,~Laplace's method), with $k=(N-1)(n-($the~dimension
of the variety on which $-\Phi$ attains its global minimum$))$, and the leading
coefficient being essentially the integral of $\phi(z;z_2,\dots,z_N)g(z_1)\dots
g(z_N)$ over that variety with respect to the corresponding Hausdorff measure;
see e.g.~\cite{Fed}, \S4~of~Chapter~II, \cite{Hrm},~Section~7.7, or~\cite{Me},
Chapter~7. This completes the proof.   \end{proof}

From a physical point of view, the sort of domains and Hilbert spaces envisaged
in the above theorem could be used to describe system having  $n$ kinematic and
$N$ internal degrees of freedom.

\section{Some related reproducing kernels} \label{DREI}
The~original motivation that led the authors to the spaces like $\bHH_h$ above
did not actually come from quantization, but~rather from an attempt to
generalize to various vector- and matrix-valued setups the multifarious
existing notions of coherent states from quantum optics (see~e.g.~\cite{AAG}).
In~our case here, these are given essentially by the ``normalized'' reproducing
kernels of the respective spaces, see~\cite{AEG}; for~instance, as~shown
\cite{AE2}, for~the spaces $\bHH_h$ of Section~\ref{ZWEI} they are just the
family $\kh_{Z,\chi}$, of elements of~$\bHH_h$, indexed by $Z\in\bOmega$ $(=\;$the
set of all normal $N\times N$ matrices) and vectors $\chi\in\CCN$, given~by
$$  \kh_{Z,\chi}(X) = \Kh(X,Z) \Kh(Z,Z)^{-1/2} \chi ,  $$
where
\begin{equation}  \Kh(X,Y) = \sum_{k=0}^\infty \frac{X^kY^{*k}} {k!h^k}
\label{dUE}  \postdisplaypenalty1000000 \end{equation}
is~the reproducing kernel of~$\bHH_h$.

Note that, despite the isomorphism from part (i) of Theorem~\ref{vBU} and the
fact that the reproducing kernel for the corresponding space $\HH_h$ is well
known to be simply~$e^{\spr{x,y}/h}$, the~reproducing kernel (\ref{dUE}) cannot
be evaluated in a closed form, since the matrices $X$ and $Y^*$ do not commute.
For~the same reason, it~is impossible to evaluate in closed form any of the
kernels from Theorem~\ref{vBU} even if the corresponding kernels for $\HH_h$
are known. All~one can do is to write them again in the form (\ref{dUE}),
only the monomials need to be replaced by some general orthonormal basis
of the space: namely, if~$\{\psi_k\}$ is~an arbitrary orthonormal basis of
$\HH_h=L^2\hol(\Omega,d\mu_h)$, then by the well-known formula of Bergman~
\cite{Be} the reproducing kernel of~$\HH_h$ is given~by
$$ \sum_{k=0}^\infty \psi_k(x) \overline{\psi_k(y)}   $$
(the~sum of the series does not depend on the choice of the orthonormal basis).
In~view of the isomorphism~(\ref{tIO}), it~therefore transpires that
\begin{equation}  \Kh(\bX,\bY) := \sum_{k=0}^\infty
\psi_k^\#(\bX) \psi_k^\#(\bY)^*   \label{dUD}  \end{equation}
is~the reproducing kernel of the space~$\bHH_h$, in~the sense that it has the
reproducing property
\begin{equation}  f(\bZ) = \intob \Kh(\bZ,\bX) f(\bX) \,d\bmu_h(\bX)
\qquad \forall \bZ\in\bOmega, \; f\in\bHH_h .   \label{tRE}  \end{equation}

In~this short section we want to call attention to some situations when the
quantization procedure from Theorem~\ref{vBU} does not apply, but~there still
exists a formula for the reproducing kernels like~(\ref{dUD}). They all arise
as Cartesian products of the spaces from Theorem~\ref{vBU} for the complex
plane~$\CC$, the~unit disc~$\DD$, and,~more generally, any one-dimensional
domains $\Omega\subset\CC$ for which the ordinary Berezin-Toeplitz quantization
works; for~simplicity of ideas, we~describe the space corresponding to $\CC^n$
(the~construction for the Cartesian product of any other $n$ spaces of the
above-mentioned type contains no additional new ideas).

The~domain in this case will consist of \emph{all} (not~just commuting)
$n$-tuples of normal matrices:
$$ \bOmega = \{\bZ=(Z_1,\dots,Z_n)\in\CCnNN: \, Z^*_jZ_j=Z_jZ^*_j
\;\forall j \}.  $$
For~the measure we take
$$ d\bmu_h(\bZ) = d\bmu_h(Z_1) \dots d\bmu_h(Z_n) ,  $$
where, abusing the notation a~little, the~$d\bmu_h$ on the right-hand side
stand for the measure (\ref{dUF}) on the set of all $N\times N$ normal matrices
from Section~\ref{ZWEI}.

Finally, we~define the space $\bHH_h$ to be the span in $L^2(\bOmega,d\bmu_h)
\otimes\CCN$~of the functions
$$ Z_1^{k_1} Z_2^{k_2} \dots Z_n^{k_n} \chi_j ,
\qquad k_1,\dots,k_n\ge0, \; j=1,\dots,N.   $$

\begin{theorem} \label{tPR} The~mapping
$$ Z_1^{k_1} Z_2^{k_2} \dots Z_n^{k_n} \chi \mapsto
z_1^{k_1} z_2^{k_2} \dots z_n^{k_n} \otimes \chi   $$
is~a unitary isomorphism of $\bHH_h$ onto~$L^2\hol(\CC^n,e^{-\|z\|^2/h}
(\pi h)^{-n}\,dz_1\dots dz_n)\otimes\CCN$. Consequently, the~reproducing
kernel of $\bHH_h$ is given~by
$$ \Kh(\bX,\bY) = \sum_{k_1,\dots,k_n=0}^\infty
\frac{X_1^{k_1}\dots X_n^{k_n} Y_n^{*k_n}\dots Y_1^{*k_1}}
{k_1!\dots k_n! h^{k_1+\dots+k_n}} .  $$   \end{theorem}

\begin{proof} Let~$Z_j=U^*_jD_jU_j$ be the spectral decomposition (\ref{tUDU})
of~$Z_j$. Then~by the definition of~$d\bmu_h$,
\begin{align*}
& \spr{Z_1^{k_1} Z_2^{k_2} \dots Z_n^{k_n} \chi,
Z_1^{j_1} Z_2^{j_2} \dots Z_n^{j_n} \eta}
= \intob \eta^* Z_n^{*j_n}\dots Z_1^{*j_1} Z_1^{k_1} Z_2^{k_2}
\dots Z_n^{k_n} \chi \; d\bmu_h(\bZ)  \\
&\qquad= (\pi h)^{-nN} \int_\CCN\dots\int_\CCN \int_\UN\dots\int_\UN
\eta^* U^*_n D^{*j_n}_n U_n \dots U^*_1D^{*j_1}_1U_1 \\
&\hskip4em\vphantom{\int} U^*_1D_1^{k_1}U_1 \dots U^*_nD_n^{k_n}U_n \chi \,
dU_1\dots dU_n \, e^{-\Tr(D^*_1D_1+\dots+D^*_nD_n)/h} \,dD_1\dots dD_n.
\end{align*}
Applying successively (\ref{tUU}) to $U_1,U_2,\dots,U_n$, we~obtain
\begin{align*}
& \frac{(\pi h)^{-nN}}N \int_\CCN\dots\int_\CCN \int_\UN\dots\int_\UN
\eta^* U^*_n D^{*j_n}_n U_n \dots U^*_2D^{*j_2}_2U_2 \Tr(D_1^{*j_1}D_1^{k_1})\\
&\qquad\vphantom{\int} U^*_2D_2^{k_2}U_2 \dots U^*_nD_n^{k_n}U_n \chi \,
dU_2\dots dU_n \, e^{-\Tr(D^*_1D_1+\dots+D^*_nD_n)/h} \,dD_1\dots dD_n  \\
&= \frac{(\pi h)^{-nN}}{N^2} \int_\CCN\dots\int_\CCN \int_\UN\dots\int_\UN
\eta^* U^*_n D^{*j_n}_n U_n \dots U^*_3D^{*j_3}_3U_3 \Tr(D_1^{*j_1}D_1^{k_1})\\
&\hskip6em\vphantom{\int}  \Tr(D_2^{*j_2}D_2^{k_2}) \; U^*_3D_3^{k_3}U_3
\dots U^*_nD_n^{k_n}U_n \chi \, dU_3\dots dU_n \\
&\hskip10em\vphantom{\int} e^{-\Tr(D^*_1D_1+\dots+D^*_nD_n)/h}
\,dD_1\dots dD_n  \\
& \hskip7em \vdots  \\
&= \frac{(\pi h)^{-nN}}{N^n} \int_\CCN\dots\int_\CCN \eta^* \Tr(D_1^{*j_1}
D_1^{k_1}) \dots \Tr(D_n^{*j_n} D_n^{k_n}) \chi \\
&\hskip10em\vphantom{\int} e^{-\Tr(D^*_1D_1+\dots+D^*_nD_n)/h}
\, dD_1 \dots dD_n   \\
&= \eta^*\chi \prod_{m=1}^n  \frac{(\pi h)^{-N}}N \int_\CCN
\Tr(D^{*j_m} D^{k_m}) \, e^{-\Tr(D^*D)/h} \,dD  \\
&= \eta^*\chi \prod_{m=1}^n  \frac{(\pi h)^{-N}}N \sum_{l=1}^N \int_\CCN
\overline{d_l^{j_m}} d_l^{k_m} \, e^{-\|d\|^2/h} \, dD  \\
&= \eta^*\chi \prod_{m=1}^n \frac1N \sum_{l=1}^N\delta_{j_m k_m} k_m!h^{k_m} \\
&= \eta^*\chi \prod_{m=1}^n \delta_{j_m k_m} k_m!h^{k_m}  \\
&= \spr{\chi,\eta} \, \spr{z_1^{k_1} z_2^{k_2} \dots z_n^{k_n},
z_1^{j_1} z_2^{j_2} \dots z_n^{j_n}}.    \end{align*}
This settles the first claim. Besides, it~shows that the function
$\Kh(\bX,\bY)$ satisfies
\begin{align*}
& \intob \Kh(\bX,\bZ) Z_1^{k_1} Z_2^{k_2}\dots Z_n^{k_n} \chi \,d\bmu_h(\bZ) \\
&\qquad= \sum_{j_1,\dots,j_n=0}^\infty \frac{X_1^{j_1}\dots X_n^{j_n}}
{k_1!\dots k_n! h^{k_1+\dots+k_n}} \intob Z_n^{*j_n}\dots Z_1^{*j_1}
Z_1^{k_1} Z_2^{k_2} \dots Z_n^{k_n} \chi \,d\bmu_h(\bZ)  \\
&\qquad= \sum_{j_1,\dots,j_n=0}^\infty \frac{X_1^{j_1}\dots X_n^{j_n}}
{k_1!\dots k_n! h^{k_1+\dots+k_n}} \; \delta_{j_1 k_1}\dots\delta_{j_n k_n}
k_1!\dots k_n! h^{k_1+\dots+k_n} \chi \\
&\qquad= X_1^{k_1}\dots X_n^{k_n},  \end{align*}
i.e.~has the reproducing property (\ref{tRE}) for functions of the form
$Z_1^{k_1}\dots Z_n^{k_n}\chi$; since the latter span all of $\bHH_h$ by
definition, the~second part of the theorem also follows.  \end{proof}

Although the reproducing kernels and the isomorphism (\ref{tIO}) work out~fine,
what breaks down is that part (ii) of Theorem~\ref{vBU}: the~Toeplitz operators
on $\bHH_h$ do not correspond, under the isomorphism above, to~the Toeplitz
operators on the Segal-Bargmann space $L^2\hol(\CC^n,e^{-\|z\|^2/h}(\pi h)^{-n}
\,dz_1\dots dz_n)$. The~reason is the noncommutativity of $Z_1,\dots,Z_n$ ---
the~reader can try to go through the beginning of the proof of part (ii) of
Theorem~\ref{vBU} to see what is happening. For~the very same reason, it~also
not possible to define spectral functions (and~much less to describe the
$U$-invariant~ones), and it is totally unclear at the moment how to achieve
anything similar to the quantization from the previous section.

We~take this occasion to remark that one lands in even greater difficulties
if one tries to deal with domains of \emph{arbitrary} (rather~than just normal)
matrices. See~Section~4 of~\cite{AE2} for~details.

On~the other hand, it~is clearly possible to use any other ordering of the
entries of $\bZ$ in Theorem~\ref{tPR} than $Z_1 \dots Z_n$ (for~instance,
$Z_n \dots Z_1$); we~omit the details.

\bigskip

\noindent{\textsc{Acknowledgement.}} Part of this work was done while
the second author was visiting the first; the~support of the Department of
Mathematics and Statistics, Concordia University, is gratefully acknowledged.

\end{document}